\documentclass[graybox, envcountchap]{svmult}

\usepackage{mathptmx}        
\usepackage{amsmath}
\usepackage{amssymb}
\usepackage{color}
\usepackage{helvet}          
\usepackage{courier}         
\usepackage{dirtree}

\usepackage{makeidx}        
\usepackage{graphicx}        
\usepackage{subfig}

\usepackage{multicol}        
\usepackage[bottom]{footmisc}

\usepackage{hyperref}        
\hypersetup{colorlinks=true,urlcolor=blue}

\usepackage[misc]{ifsym}

\usepackage{aas_macros} 

\makeindex             

\begin{document}

\title{Astrophysical X-Ray Polarization}
\author{Philip Kaaret and Brian D. Ramsey}

\institute{Philip Kaaret (\Letter) \at NASA Marshall Space Flight Center, Huntsville, AL 35812, USA, \email{philip.kaaret@nasa.gov}
\and Brian D. Ramsey (\Letter) \at NASA Marshall Space Flight Center (Retired), Huntsville, AL 35812, USA, \email{brian.ramsey@nasa.gov}}

%
%
\maketitle

\abstract{X-ray polarimetry is now providing a new way to look at the high energy sky. The addition of two observables, polarization fraction and angle, reveals crucial new information on the structure of accretion flows and magnetic fields in astrophysical systems. Here, we review the basic physical processes that produce polarized X-rays in astrophysical contexts. Then, we briefly describe the physical processes used to measure X-ray polarization and the detectors that have been flown or are under construction.}


\section{Introduction}
\label{sec:intro}

X-ray polarization conveys information on the geometry of high-energy astrophysical sources and the emission processes at work. The polarization signal may reflect the physical distribution of the emitting matter, the arrangement of the magnetic or radiation fields, or even the geometry of space-time. The information that can be gleaned via X-ray polarimetry provides insights unavailable from other observational techniques, often with significant implications for our understanding.

In the past decade, X-ray polarization has blossomed from confident detection of a single source, the spatial average from the Crab nebula \cite{Novick1972,Weisskopf1978}, to detections or model-constraining upper limits on dozens of sources distributed across many different classes of astrophysical objects. The largest contribution has come from the launch and successful operation of the Imaging X-ray Polarimetry Explorer (IXPE, see Chapter~14). As of this writing over 350 IXPE-related refereed papers have been published with over 6000 citations. These new X-ray polarization results have provided fundamentally new information and often challenged our pre-existing views.

This chapter provides an introduction for those previously unfamiliar with X-ray polarimetry. After a brief discussion of how to describe polarized X-rays (Section~\ref{sec:stokes}), we introduce the primary physical mechanisms that generate polarized emission in astrophysical objects (Section~\ref{sec:emission}). Then, we describe the physical processes used to measure X-ray polarization and detectors utilizing those processes (Section~\ref{sec:detection}).

\section{Description of Polarized X-Rays}
\label{sec:stokes}

X-rays are discrete packets of electric and magnetic fields oriented transverse to the direction of motion. Polarization describes the configuration of the fields. Since the electric and magnetic fields are interrelated by Maxwell's equations, it is sufficient to specify only the configuration of the electric field. The electric field of an X-ray propagating along the $z$-axis can be described as
\begin{eqnarray}
\vec{E} = \hat{x} E_{X} + \hat{y} E_{Y} = 
          \hat{x} E_{0X} \cos(kz - \omega t) +
          \hat{y} E_{0Y} \cos(kz - \omega t + \xi).
\label{eq:efield}
\end{eqnarray}

\noindent where the angular frequency ($\omega = 2 \pi E / h$) and the wavenumber ($k = \omega / c = 2 \pi E / h c$) are both determined by the photon energy ($E$) \cite{KaaretWSPC}. For a photon moving in an arbitrary direction, the wavevector ($\vec{k}$) is defined with magnitude equal to the $k$ and direction along the photon momentum.

The polarization is determined by the phase offset ($\xi$) and the ratio of amplitudes between the two electric field components. When $\xi = n \pi$, where $n$ is an integer, then $E_{X}$ and $E_{Y}$ are exactly in phase (or antiphase) and their amplitudes are always proportional. The photon is then described as linearly polarized. If $\xi \ne n \pi$, then the fields rotate as the photon propagates. This is elliptical polarization. Circular polarization is the special case of elliptical polarization with $E_{0X} = E_{0Y}$.

The net polarization of a set of photons from a source are fully characterized by the four Stokes parameters:
\begin{eqnarray}
I = \langle E_{0X}^2 \rangle + \langle E_{0Y}^2 \rangle \\
Q = \langle E_{0X}^2 \rangle - \langle E_{0Y}^2 \rangle \\
U = \langle 2 E_{0X} E_{0Y} \cos \xi \rangle \\
V = \langle 2 E_{0X} E_{0Y} \sin \xi \rangle 
\label{eq:stokes}
\end{eqnarray}

\noindent where the averages are over the photons detected from the source and may be calculated in specific bins of energy, time, pulse or orbital phase, etc. Stokes $I$ is the total intensity. This is the same intensity as measured by instruments insensitive to polarization. Stokes $Q$ and $U$ describe linear polarization. Stokes $V$ describes elliptical polarization.  Since all X-ray polarimeters flown to date are sensitive only to linear polarization, we will not consider elliptical polarization further.

\begin{figure}[tb]
\centerline{
\includegraphics[width=0.6\textwidth, keepaspectratio=true]{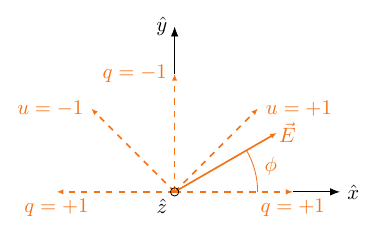}}
\caption{Polarization `ellipse' diagram for linearly polarized radiation. The figure shows a view of the photon electric field looking into the beam; the z-axis is out of the page. The electric field ($\vec{E}$) is shown for polarization angle (EVPA) $\phi = 30^{\circ}$. Also shown are the electric field orientations for $q = +1 \rightarrow \phi = 0^{\circ}$, $u = +1 \rightarrow \phi = 45^{\circ}$, $q = -1 \rightarrow \phi = 90^{\circ}$, $u = -1 \rightarrow \phi = 135^{\circ}$.} \label{fig:polell}
\end{figure}

The normalized Stokes parameters for linear polarization are $q = Q/I$ and $u = U/I$. These range between $+1$ and $-1$. For linear polarization, the polarization fraction (also referred to as the polarization degree) is $\Pi = \sqrt{Q^2 + U^2}/I = \sqrt{q^2 + u^2}$ and is often quoted as a percentage. The electric vector position angle (EVPA), also referred to as the polarization angle, is $\phi =  \tfrac{1}{2} \arctan(U/Q) = \tfrac{1}{2} \arctan2(U, Q)$. EVPA is defined modulo a 180$^{\circ}$ rotation. The electric field configurations for various values of $q$ and $u$ are shown in Fig.~\ref{fig:polell}. Note that a change in sign of $q$ or $u$ rotates $\vec{E}$ by $90^{\circ}$.

\section{Emission of Polarized X-Rays}
\label{sec:emission}

Interpretation of the X-ray polarization measured from an astrophysical source requires an understanding of physical processes that produce the X-radiation. Here, we provide brief descriptions of the some physical processes that produce polarized X-rays in high-energy astrophysical objects.

\subsection{Thomson/Compton scattering}
\label{sec:compton}

Scattering between photons and electrons is a key process in the generation of the X-rays that we observe from many astrophysical systems. For example, active galactic nuclei and accreting stellar‑mass black holes produce X-rays via scattering in accretion disks and hot coronae. The polarization of the observed X-rays is determined by the geometry and other properties of the scattering medium. Hence, X-ray polarimetry can provide information on the structure of accretion flows in astrophysical objects.

Scattering between photons and free electrons is labeled as Thomson, Compton, or inverse-Compton. Thomson scattering is the elastic limit where the photon energy ($E$) is unchanged by the scatter and occurs in the non-relativistic limit when $E \ll m_e c^2$, where $m_e$ is the electron mass. Compton scattering is the inelastic case where energy is transferred between the photon and electron. The term `inverse-Compton scattering' is often used when energy is transferred from electrons to photons. 

\begin{figure}[tb]
\includegraphics[width=\textwidth, keepaspectratio=true]{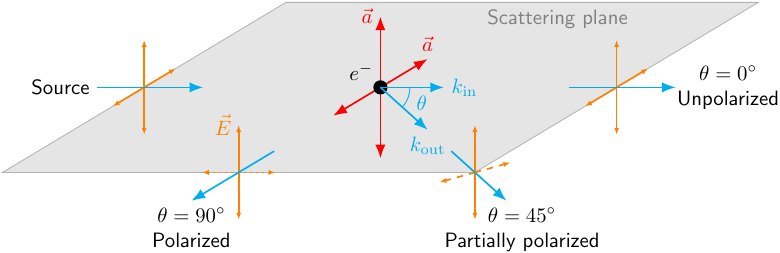}
\caption{Geometry of Thomson Scattering. The source/incoming photon (left) interacts with an electron (center) producing a scattered/outgoing photon (right, lower right, and near). Blue vectors indicate photon momenta while orange vectors indicate photon electric fields ($\vec{E}$). Red vectors ($\vec{a}$) indicate acceleration of the electron (black circle $e^-$). The scattering angle ($\theta$) is defined by the wavevectors of the incoming ($k_{\rm in}$) and outgoing ($k_{\rm out}$) photons. The scattering plane (gray) contains both $k_{\rm in}$ and $k_{\rm out}$.} \label{fig:unpolthomson}
\end{figure}

We begin with the simplest case, that of Thomson scattering, shown in Fig.~\ref{fig:unpolthomson}. The electric field of the incident photon causes the electron  ($e^{-}$) to oscillate. The resulting acceleration ($\vec{a}$) of the electron produces radiation in the form an outgoing photon ($k_{\rm out}$). The scattering angle ($\theta$) is defined by the wavevector of the incoming ($\gamma_{\rm in}$) and outgoing ($k_{\rm out}$) photons. The scattering plane contains both $k_{\rm in}$ and $k_{\rm out}$. Photons from an unpolarized source have electric fields in the plane perpendicular to $k_{\rm in}$ and are able to accelerate ($\vec{a}$) the electron in the same plane. The electric field of the photon $k_{\rm out}$ produced by acceleration of the electron, must therefore also lie in the plane perpendicular to $k_{\rm in}$.

Thomson scattering in the forward direction ($\theta = 0^{\circ}$) produces unpolarized outgoing radiation for an unpolarized input source. In this case, the system is symmetric around the common axis of $k_{\rm in}$ and $k_{\rm out}$ and all electron acceleration directions radiate electric fields in directions accessible for the outgoing photon. In contrast, for Thomson scattering in the perpendicular direction ($\theta = 90^{\circ}$), only acceleration in the direction perpendicular to the scattering plane produces electric fields accessible to the outgoing photons. This leads to 100\% polarized outgoing radiation with the electric field perpendicular to the scattering plane. More generally, the degree of polarization for scattering of unpolarized source radiation varies as a function of scattering angle as

\begin{equation}
\Pi = \frac{1 - \cos^2 \theta}{1 + cos^2 \theta}
\label{eq:polthomson}
\end{equation}

\noindent and the net electric field is oriented perpendicular to the scattering plane \cite{Rybicki1986}. In the Thomson approximation, the scattering cross section including polarization dependence are independent of energy. At higher energies, above a few tens of keV, inelastic and relativistic effects described by the Klien-Nishina cross section become important.

The X-ray emission observed from astrophysical sources where the central engine is obscured can be dominated by X-rays that have undergone a single scattering through a large angle. High polarization is expected in such cases. Examples include obscured active galactic nuclei (AGN of Type 2) and obscured stellar-mass black holes such as Cyg~X-3. High polarization fractions, in the range of 10\% to 30\% are observed from such systems and confirm their obscured nature \cite{Ursini2023,Veledina2024}. The measured polarization angle can be used to constrain the source geometry projected on the sky, while the polarization degree, via Eq.~\ref{eq:polthomson}, can be used to constrain the geometry along the line of sign. A unique example is detection of polarization from molecular clouds near the Galactic Center thought to be reflected radiation from an (unobserved) past outburst from Sgr A* \cite{Marin2023}.

The discussion above is for a single scattering. Multiple scattering is often important in astrophysical structures such as accretion disks and coronae. When the number of scatterings is large, photons 'forget' about the properties of the primary sources \cite{Sunyaev1985}. Instead, the observed polarization is determined by the properties of the scattering medium, notably the optical depth.

The seminal result was by Chandrasekhar who studied radiation transfer via Thomson scattering in a plane parallel atmosphere \cite{Chandrasekhar1946}. The results are directly applicable to optically thick and geometrically thin disks as found in many accreting systems. The polarization of the outgoing photons is linear with the electric field in the disk plane. The polarization fraction is zero for a face-on disk, increasing to a maximum of 11.7\% for a disk viewed edge on. The polarization angle lies along the disk major axis as projected on the sky.

Many of the X-rays observed from astrophysical sources are produced in hot coronae where photons gain energy by repeatedly scattering on hot electrons, a process known as `thermal Comptonization'. First treatments of the process used the Thomson approximation applied to a disk geometry and found that emitted polarization depends only on the optical depth of the scattering medium and not on the photon energy \cite{Sunyaev1985}. Remarkably, the polarization changes from parallel to the scattering surface for high optical depth to perpendicular to the surface for low optical depth. For high optical depth, photons diffuse towards the surface and their momenta are peaked parallel to the surface normal. Thus, the electric fields of escaping photons tend to lie parallel to the surface. For low optical depth, the photons most likely to be scattered are those travelling parallel to the plane of the scattering material. The electric field is, thus, predominantly perpendicular to the plane \cite{Angel1969}.

While these simple models can provide insight, more complex models including both more accurate physics and more complex accretion structures are needed to accurately represent astrophysical systems. Use of the relativistic Klein–Nishina cross section and the Compton scattering matrix is important even at electron temperatures or photon energies as low as 50~keV \cite{Poutanen1996}. Reflection, or scattering of photons produced in one part of the accretion flow off another part, can produce significant polarization. Reflection can arise as coronal photons scattering by the accretion disk or, in strong gravitational fields, as `returning radiation' - disk photons that are bent along geodesics and strike the disk \cite{Schnittman2009}. General relativistic frame dragging, the Lense-Thirring effect, causes the spacetime around rapidly spinning black holes to rotate; this can lead to rotation of the polarization angle with energy \cite{Connors1980}. More complex structures, such as centrally illuminated disks or disks with winds \cite{Matt1993}, can have significant polarization signatures. IXPE has revealed higher polarization than expected in a number of sources where the emission is dominated by scattering and this arena is ripe for exploration as discussed in Part~3 of this volume.

\subsection{Synchrotron radiation}

Astrophysical systems that host relativistic electron populations moving in magnetic fields necessarily produce synchrotron emission. Such environments arise in pulsar wind nebulae and supernova remnants (see Chapter~6), and in jets from active galactic nuclei and stellar-mass black holes. X-ray polarization provides information on the magnetic field configuration in these source furnishing insight into the acceleration mechanism. X-ray polarization can also help distinguish between different emission mechanisms and help determine the jet composition.

\begin{figure}
\centerline{
\includegraphics[width=0.7\textwidth, keepaspectratio=true]{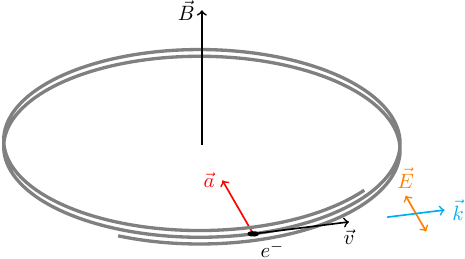}}
\caption{Geometry of Synchroton Radation. An electron (black circle $e^-$) moves in a helical path (gray) in the plane perpendicular to the magnetic field (black vector $\vec{B}$). The electron's acceleration (orange vector $\vec{a}$) produces a photon (blue wavevector $\vec{k}$) with the electric field (orange vector) in the plane of motion. The radiation is beamed in the  electron's direction of motion (black vector $\vec{v}$).}
\label{fig:synchro}
\end{figure}

The basic physics is that electrons accelerate as they move in move in a helical path in the magnetic field and then the electrons' acceleration produces radiation. The geometry is shown in Fig.~\ref{fig:synchro}. An electron ($e^{-}$) spirals in a helical pattern (gray) perpendicular to the magnetic field ($\vec{B}$). The electron's acceleration ($\vec{a}$), and therefore, the electric field of the emitted photon with wavevector ($\vec{k}$) lies in the plane perpendicular to the magnetic field. For relativistic electrons, the radiation is beamed in the  electron's direction of motion ($\vec{v}$). The resulting radiation is intrinsically linearly polarized, with its electric field perpendicular to the magnetic field. 

For an isotropic population of electrons with a power-law distribution of energies with an electron spectral index of $p$ in a uniform magnetic field, the maximum polarization for synchrotron radiation is $\Pi = (p+1)/(p+7/3)$ \cite{Rybicki1986}. The electron spectral index is related to the observed photon index ($\Gamma$) as $p = 2\Gamma - 1$. In astrophysical contexts, the magnetic field is typically not uniform. If the field is decomposed into a uniform component and a random component, then the observed polarization is reduced by a factor approximately equal to the ratio of the energy in the uniform component to the energy in the total field \cite{Burn1966}.

\subsection{Photon propagation in strong magnetic fields}

The theory of Quantum Electrodynamics (QED) predicts that strong magnetic fields render the vacuum birefrigent, meaning the propagation of photons depends on their polarization \cite{Heisenberg1936}. A classical photon does not interact with a magnetic field in vacuum. However, in QED there is a non-zero probability for a photon to (briefly) convert to a virtual electron-positron pair that will interact with the magnetic field. This couples the photon to the magnetic field. The critical magnetic field strength for QED effects is the strength at which the electron cyclotron energy equals its rest mass, $B_Q = 4.4 \times 10^{13}$~G. Significant birefringence occurs for fields near or exceeding the critical field strength.

Polarization of the photons is described in terms of two modes defined by the plane formed by the photon wave vector, $\vec{k}$, and the neutron star magnetic field, $\vec{B}$. The ordinary mode (O-mode) is polarized parallel to the $\vec{k} - \vec{B}$ plane, while the extraordinary mode (X-mode) is polarized perpendicular to the $\vec{k} - \vec{B}$ plane. For X-rays propagating in fields with strengths near $B_Q$, the X-mode opacity is greatly suppressed relative to the O-mode opacity so the radiation emerging from the atmosphere is primarily in the X-mode and, thus, strongly polarized.

Vacuum birefringence couples the polarization modes to the magnetic field so that as an X-ray propagates though the magnetosphere, its EVPA follows the direction of the magnetic field out to large radii \cite{Adler1971,Heyl1997}. The result is that the observed EVPA is correlated with the direction of the magnetic field far from the neutron star, even for photons originating from different points on the neutron star with different surface field directions \cite{Heyl2002}. This can greatly enhance the observed polarization signal. 

High magnetic field strengths, nearing or even exceeding the critical field strength, are found in neutron stars. Accreting X-ray pulsars contain neutron stars with magnetic fields of $10^{12} - 10^{13}$~G. The effects of birefringence are important for the radiation transfer in these systems, see Chapter~11. Some of the highest X-ray polarization fractions observed are from magnetars, neutron stars with ultra-strong magnetic fields perhaps as high as $10^{15}$~G. Birefringence effects are of order unity in these systems, hence, they provide ideal natural laboratories to search for vacuum birefringence, see Chapter~6.


\section{Detection of Polarized X-Rays}
\label{sec:detection}

\begin{figure}[tb]
\centering
\includegraphics[width=3.25in, keepaspectratio=true]{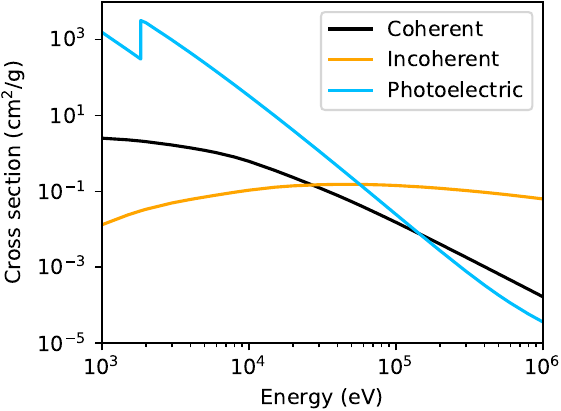}
\caption{Photon interaction cross sections versus energy for silicon.}
\label{fig:cross}
\end{figure}

Techniques for detecting polarized X-rays vary with energy as the interaction process of photons with the detection medium is energy dependent. Figure \ref{fig:cross}, generated for silicon as a typical low-atomic-number detector material, shows that at low energies the photoelectric effect dominates, followed by coherent (Thomson) scattering, whereas at higher energies, above about 50~keV, incoherent (Compton) scattering becomes the dominant interaction mechanism. For lighter materials, scattering will become dominant at lower energies. For beryllium, for example, scattering will be dominant at just over 10~keV.

In the photoelectric effect, an incident X ray interacts with an electron in the absorber, typically from the inner-most shell energetically possible, ejecting a photoelectron and leaving a vacancy in the atom which is later filled through emission of radiation. The whole of the energy of the incident photon ($E_p$) is used in this process, resulting in a photoelectron of energy $E_p$ minus the binding energy of the electron and any additional energy expended in exciting the atom.

In scattering, the incident photon scatters with an orbiting electron (typically from an outer shell). For incoherent scattering, also called Compton scattering, the deflected photon loses energy by an amount dependent on the angle through which it has scattered. This energy is imparted to the electron, a so-called Compton electron, which is emitted from the atom. For coherent, or Thomson, scattering there is essentially no loss of energy, no electron is emitted, and the deflected photon continues in a new direction but with the same energy as before.

The photoelectric and the scattering processes all produce signals which contain information about the electric vector of the interacting photon; the former via the emission direction of the photoelectron and the latter via the angular dependence of the scattered photons. Designing a sensitive instrument to exploit this information is the key to efficient detection of polarized X rays. The modulation factor ($\mu$) is a measure of the response of the polarimeter to 100$\%$ polarized X-rays. An ideal polarimeter has $\mu = 1$, while an instrument not sensitive to polarization has $\mu = 0$.

The sensitivity of a polarimeter is commonly quantified using the minimum detectable polarization (MDP) which is the smallest polarization degree that can be measured within a given integration time ($T$) for specified source ($R_s$) and background  ($R_b$) counting rates. For historical reasons, MDP is usually specified at the 99$\%$ confidence level for which it is

\begin{equation}\label{eq:MDP}
    MDP = \frac{4.29}{\mu} \times 
        \frac{\sqrt{(R_{s} + R_{b})\times t}}{R_{s}\times t}
\end{equation}

\noindent It can be seen that the MDP scales with the inverse of the modulation factor, but only as the inverse of the $\textit{square root}$ of the source counts. Thus, optimizing the former is imperative. Also, the above equation shows that to achieve percentage-level polarization sensitivity ($MDP = 1\%$), even in the absence of background, requires $\sim$ 1.8 x 10$^{5}$ photons if the modulation factor ($\mu$) is 1, and $\sim$ 2.0 x 10$^{6}$ photons if $\mu$ is just 0.3. Given the low fluxes of most cosmic X-ray sources, this typically means long integration times for sensitive studies, pointing to the need for dedicated missions.

\subsection{Measurement Techniques}


\subsubsection{Photoelectric Polarimeters}

\begin{figure}[hbt]
\centering
\includegraphics[width=0.9\textwidth, keepaspectratio=true]{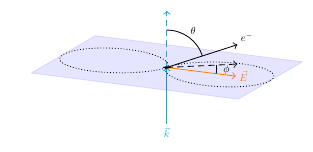}
\caption{Angular distribution of K shell photoelectrons from a polarized photon beam. The photon (blue vector $\vec{k}$) is incident from the bottom of the page and interacts with an atom (black circle) ejecting a photoelectron (black arrow $e^-$). The dashed black arrow is the projection of the photoelectron direction on the the plane perpendicular to the photon wavevector. The photoelectron azimuthal angle ($\phi$) is measured relative to the photon electric field and the photoelectron polar angle ($\theta$) measured relative to the photon wavevector. The dotted curves shown the $\cos^2\phi$ dependence of the cross section for $\theta = 90^{\circ}$} \label{fig:Kshell} 
\end{figure}

The photoelectric polarimeter combines detection and polarization analysis and is typically more efficient than a scattering polarimeter. At low energies, photoelectrons tend to be ejected in the direction of the electric vector of the absorbed photon, at right angles to the direction of incidence. The angular dependence of the cross section is

\begin{equation}
\frac{d\sigma}{d\omega} \propto
  \frac{\sin^2\theta \cos^2\phi}{(1-\beta \cos\theta)^4}
\end{equation}

\noindent where the angles $\theta$ and $\phi$ are as defined in Figure \ref{fig:Kshell} and $\beta = v/c$ where $v$ is the photoelectron velocity.

The challenge in implementing a photoelectric polarimeter is to measure the photoelectron track, produced as the photoelectron comes to rest in the absorbing medium producing a trail of ionization. It is the initial direction of this track, as the photoelectrons exits the absorbing atom,  that contains the vital polarization information, but this portion of the tracks contains the lowest ionization density. Also, at low energies these photoelectrons have very short range, particularly in solids. For example, at 10 keV, the range of an electron is just 1 µm in silicon, much smaller than the typical pixel size in a CCD. However, in light gases, the ranges are $\sim 1000 \times$ larger, and this makes imaging the photoelectron track feasible. The challenge now is to optimize the gas filling and the readout system. The dichotomy is clear: high fill-gas pressure and atomic number are needed for high X-ray absorption efficiency, whereas low atomic number gases at low pressure provide the longest, easily measurable photoelectron tracks. In addition, increasing the depth of the absorption region to improve efficiency causes more diffusion of the electron cloud as it drifts to the sensing region, resulting in a loss of track information. A further complication is that in the photoelectric process, the absorbing atom is left with an inner shell vacancy which is filled through the emission of further radiation (Auger electron or X-ray). This results in additional ionization which, if too large, can further complicate the determination of the initial photoelectron track direction. Thus, ideally, a gas should be chosen so that it has a very low K shell energy compared with the energy range of interest.

\subsubsection{Bragg Polarimeters}

In Bragg refraction, the reflection of X rays from successive crystal planes leads to constructive interference, providing a very high reflectivity if the Bragg condition is met. This occurs when the path length between X rays reflected off successive planes, $2 d \sin (\theta)$, where $\theta$ is the incident angle relative to the lattice plane and $d$ is the lattice spacing, is an exact multiple of the incident X-ray’s wavelength. Near 100$\%$ reflectivity can be obtained at incident angles near $45^\circ$. Since the reflectivity is maximum when the X-ray has its electric vector normal to the plane and is effectively zero for X-rays with electric vector parallel to the lattice plane, Bragg polarimeters can provide a high modulation factor.

\begin{figure}[tb]
\centering
\includegraphics[width=2.5in, height=2.83in, keepaspectratio=true]{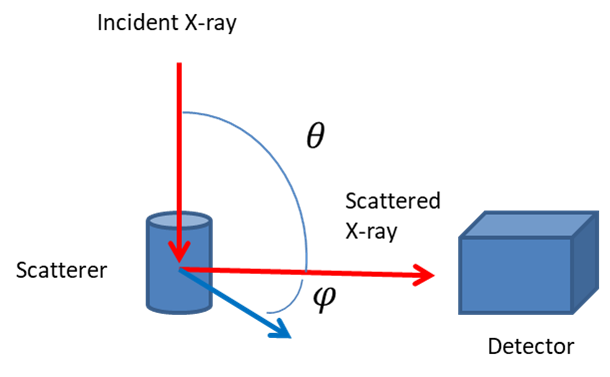}
\caption{Thomson polarimeter. An X-ray is scattered by a target and subsequently detected. The azimuthal angle ($\phi$) distribution of detected photons provides a measure of the linear polarization of the incoming beam. The modulation factor is largest for scattering angles ($\theta$) near $90^{\circ}$.} \label{fig:Thomson} 
\end{figure}

\subsubsection{Thomson Polarimeters}

The principal operation of a Thomson polarimeter is shown in Figure \ref{fig:Thomson}. An incident X-ray scatters off a target and is then absorbed in an adjacent detector. The detector records the azimuthal distribution of the scattered photons either by rotation around the X-ray axis, or by surrounding the scatterer with a series of detectors. The system is optimized for scattering angles of $90^{\circ}$ which maximizes the polarization response, see section~\ref{sec:compton} on Thomson/Compton scattering.

To operate efficiently, the scatterer must maximize the ratio of scattered to absorbed photons, and this typically means choosing the lightest possible elements, either lithium or beryllium. The detector, on the other hand, must be chosen to photoelectrically absorb the scattered photon, for total energy deposition.

\begin{figure}[tb]
\centering
\includegraphics[width=0.9\textwidth, keepaspectratio=true]{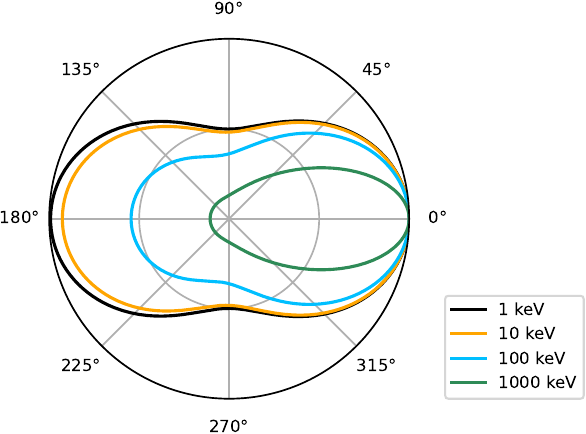}
\caption{Klein-Nishina cross section for Compton scattering interactions versus scattering angle. Cross sections are normalized to the square of the classical electron radius.} \label{fig:Klein} 
\end{figure}

\subsubsection{Compton Polarimeters}
\label{sec:comptonpol}

In the hard-x-ray region the Compton (incoherent) scattering process dominates. As polarized X-rays preferentially scatter normal to the polarization plane, the azimuthal distribution of the scattered photons contains the information on the polarization of the incident radiation. Scattering at 90° provides the biggest modulation factor for polarized X rays. But, as shown in Figure \ref{fig:Klein}, the Klein-Nishina differential cross section for Compton-scattered photons shows the smallest value at 90° for hard-X-ray energies with scattering preferentially in the forward or backward directions, and this reduces the modulation factor for scattering instruments.

\begin{figure}[tb]
\centering
\includegraphics[width=3.5in, keepaspectratio=true]{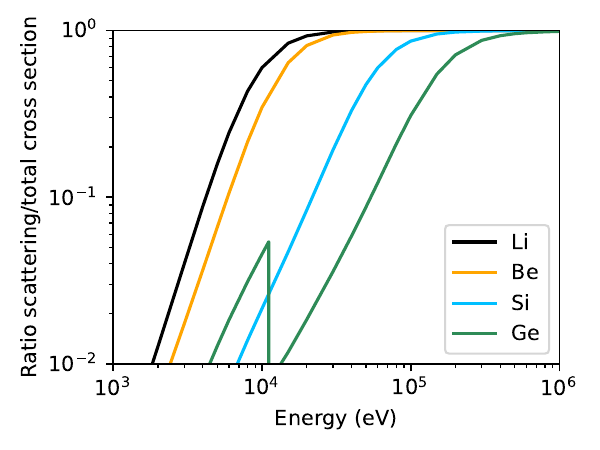}
\caption{Ratio of scattering to total cross section as a function of energy for various materials.} \label{fig:scat} \end{figure}

A Compton polarimeter needs a scatterer and an absorber as for the Thomson polarimeter, see Figure \ref{fig:Thomson}. The device could be of a pixelated design, with an array of identical channels acting as both scatterer and absorber or an array of low Z scatterers surrounded by high-Z absorbers \cite{Costa2024}. In the former configuration, coincident signals, one from the scattering pixel and one from the absorbing pixel, can be used to reduce the significant background inherent in all large-area detectors. Reconstruction of the Compton interaction allows for projection back onto the sky and makes coarse imaging possible as energy is deposited at the initial interaction site and thus scattering angles can be inferred, unlike in Thomson polarimeters. Various gamma ray instruments have used this pixelated/segmented approach for polarimetry (e.g., PoGO \cite{Friis2018}) and for gamma-ray burst polarimetry (e.g., GRAPE \cite{McConnell2009}). In addition, pixelated instruments not designed as polarimeters have been used to perform polarization measurements -- a measurement of the polarization of the Crab pulsar and nebula was obtained with the IBIS Telescope on the INTEGRAL mission \cite{Forot2008}). See \cite{Chattopadhyay2021} for a comprehensive review of Hard-X-Ray polarimetry.

An alternative approach places a very-low-Z inactive scatterer at the focus of hard-X-ray optics. The scatterer is surrounded by an array of absorbers which, because of the compact focal plane design, can be significantly smaller than in the fully-pixelated designs described above. This helps with background reduction which is necessary as the very-low-Z focal-plane scatters do not provide an active coincidence signal. 

The choice of materials is dependent on the energy range of operation. An ideal scatterer would have a large cross section for Compton interactions (compared with the photoelectric cross section) to give a high probability of scattering but a low probability of re-absorption in the scatterer, while an ideal absorber would have just the opposite – that is, a high photoelectric to Compton cross section ratio to maximize the chance of total absorption. Figure \ref{fig:scat} shows the ratio of scattering to total cross sections for various potential scatterer materials as a function of energy.

\subsection{X-Ray Polarimeters}

\subsubsection{Early measurements (1970s)}

The first statistically-significant measurements of polarization of a cosmic X-ray source were done in 1971 by a group from Columbia University, flying a sounding rocket payload equipped with both Thomson and Bragg crystal polarimeters \cite{Novick1972}.  The forward end of the payload was equipped with an array of lithium scattering blocks surrounded by absorbing detectors while aft were four Bragg crystal polarimeters, each utilizing graphite mosaic crystals oriented at 45 to the incident flux. The use of mosaic crystals, ‘imperfect’ crystals with a small range of crystal plane angles, provides a greater throughput (2-3.2~keV) than a pure crystal which only operates over a very tiny energy band (at 2.6~keV) satisfying the Bragg condition at one incident angle. Combining the data from both the Thomson and the Bragg polarimeters lead to a measurement of $15.4\% \pm 5.2\%$ polarization for the Crab Nebula.  More sensitive, confirming measurements were carried out in 1976 and 1977 using the graphite crystal polarimeter on board the Orbiting Solar Observatory (OSO-8). Here, the crystals were curved to focus the reflected flux onto small detectors thereby reducing the background and improving sensitivity. This led to a measurement of the polarization of the Crab Nebula, with pulsar contribution removed, of $19.2\% \pm 1.0\%$ near 2.6~keV \cite{Weisskopf1978}.

\subsubsection{The Dark Ages (1980s-2000s)}

These early measurements were limited by collecting area and integration time, and so were restricted to the brightest X-ray sources.  The Stellar X-ray Polarimeter (SXRP) planned for flight aboard the Soviet Spectrum-Rontgen-Gamma (SRG) mission, attempted to alleviate this limitation by placing a polarimeter at the focus of an X-ray telescope. The polarimeter featured both a graphite crystal Bragg polarimeter and below it a Thomson scattering polarimeter consisting of a lithium block housed in a beryllium vessel. Surrounding the whole assembly was a 4-sided gas-filled proportional counter \cite{Kaaret1990}. By separating the flux collection, via X-ray telescope, from the polarimeter, a large collecting area could be obtained without a commensurate increase in background. The effective source-flux collection area was $\sim$1500~cm$^{2}$, whereas the effective detector area was just $\sim$50~cm$^{2}$ \cite{Costa2024}, so a significant increase in signal to noise was obtained. The use of an optic with degree-level acceptance angles, also limited potentially-contaminating flux from nearby sources. Unfortunately, being one of several instruments in the focal plane meant that integration time was going to be limited for the SXRP instrument. Sadly, after the SXRP flight instrument had been completed, the collapse of the Soviet Union in 1989 led to the cancellation of the program.

\begin{figure}[tbh]
\includegraphics[width=4.0in, height=2.83in, keepaspectratio=true]{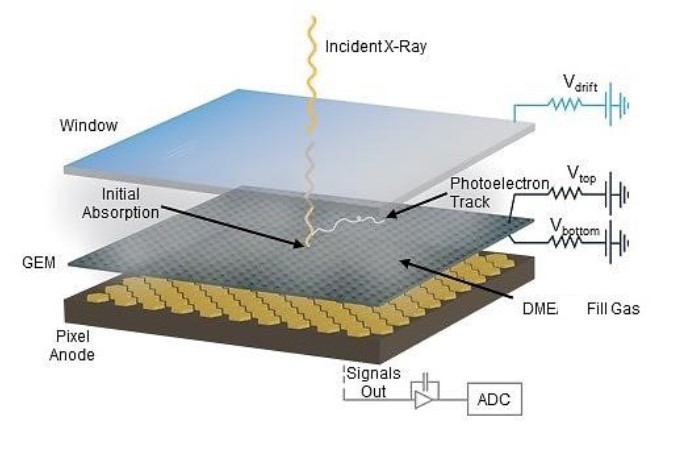}
\caption{IXPE Gas Pixel Detector (GPD) schematic. Credit: INAF-INFN.} \label{fig:GPD} \centering
\end{figure}

\subsubsection{Gas Pixel Detectors}

The renaissance that ended the dark ages was development of photoelectric polarimetry as articulated in a key paper by Costa et al.\ \cite{Costa2001}. They constructed an X-ray polarimeter consisting of a gas detection/drift region, a gas electron multiplier (GEM), and a multi-pixel, two-dimensional electronic read-out which would later be named the Gas Pixel Detector (GPD). GPDs offer a significant improvement in sensitivity relative to classical scattering polarimeters due to the intrinsic efficiency of the photoelectric process combining both polarization analysis (via the photoelectron track initial direction) and photon detection (via the photoelectron track). An additional benefit of GPDs is two-dimensional imaging. When placed at the focal plane of an X-ray telescope, GPDs offer the possibility of resolving the focal spot thereby providing a very large reduction in detector background and a commensurate improvement in overall signal to noise. Further, such a system enables spatially-resolved polarimetry, providing a wealth of information for extended sources.

GPDs were proposed for multiple missions beginning shortly after their development. The first GPD flown was PolarLight in 2018 on a Chinese cubesat mission without a focusing optic \cite{Feng2019}. The next were the GPDs on IXPE flown in 2021 with focusing optics, thus providing greatly enhanced sensitivity. Figure \ref{fig:GPD} shows a schematic of the IXPE GPD, see Chapter~14. X-rays enter through a beryllium window and interact with the fill gas, producing a photoelectron track. The electrons in this track then drift, under the influence of an electric field, towards a gas electron multiplier, where each electron in the track is then ‘amplified’, by a factor of $\sim 100 \times$ to produce a signal on the pixel anode readout array. The latter is a custom ASIC with 300 x 352 hexagonal pixels on a 50 µm pitch. Each pixel contains an independent readout chain. 

Optimization of the fill gas composition, pressure and depth factors in the energy response of the X-ray mirror module, at whose focus the detector sits, as well as the energy spectrum of typical targets that IXPE observes. The detector design parameters were varied to determine the combination giving the highest polarization sensitivity. The flight configuration chosen was a fill gas of dimethyl ether (CH$_3$OCH$_3$), known to have very low diffusion, at 800 mbar pressure and an absorption depth of 10 mm. This combination results in photoelectron tracks with length $\sim0.5$~mm at 5.9~keV, which is well sampled by the 50-µm-pitch readout. While the quantum efficiency of this configuration is modest (15$\%$ at 3 keV), it emphasizes the modulation factor ($\sim$ 0.4 at 3 keV), on which the polarization sensitivity is strongly dependent (see eqn. \ref{eq:MDP}).

\begin{figure}[ht]
\includegraphics[width=\textwidth, keepaspectratio=true]{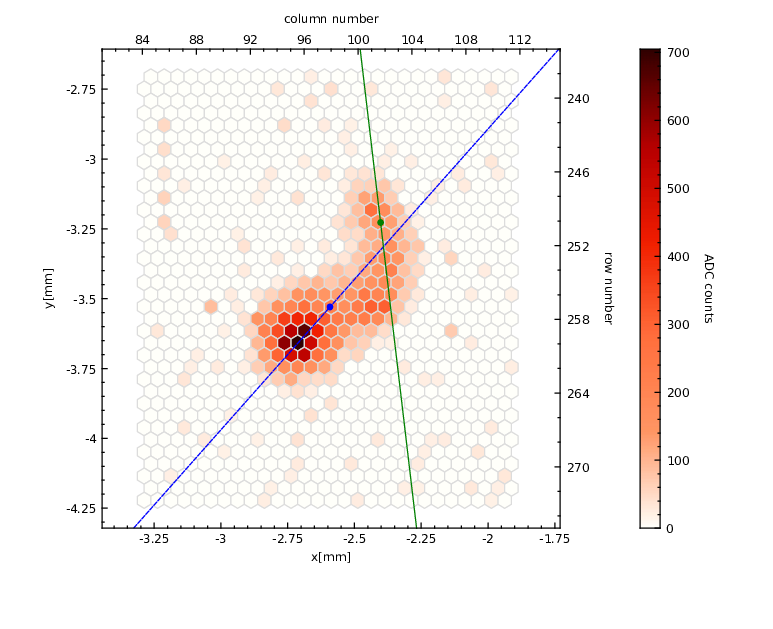}
\caption{Gas pixel detector image of a photoelectron track from a 5.9~keV photon. The green dot indicates the calculated initial photon interaction site and the green line the calculated photoelectron emission direction. Credit: INAF-INFN.} \label{fig:Track} \centering
\end{figure}

Analysis of the tracks presents something of a challenge. The desired information is contained at the very beginning of the track, but the photoelectron can undergo large angle scattering and most of the charge is deposited at the end of the track. Figure \ref{fig:Track} shows an actual track from the interaction of a 5.9 keV photon. The standard analysis for IXPE, see Chapters~2 and 3, starts by using a moments analysis to determine a principal axis and a center of gravity (blue line and blue dot in the figure) then de-weights the region of high charge density (recognized as the end of the track) and recalculates via a second moment analysis to derive the initial interaction point (green dot) and photoelectron emission direction. This process works well at higher energies where tracks are well defined, as in Figure \ref{fig:Track}, but less so at lower energies where the very short track lengths and diffusion in the detector gas makes discerning the initial photoelectron emission direction a challenge. This results in a lower energy limit near 2~keV for the IXPE detectors. More advanced processing, such as neural network based algorithms, can improve track reconstruction \cite{Kitaguchi2019,Peirson2021}.

The next instrument to use a GPD will likely be the Polarimetry Focusing Array (PFA) aboard the enhanced X-ray Timing and Polarimetry  (eXTP) mission, a Chinese-led program planned for launch later this decade. PFA will feature gas pixel detectors, similar to IXPE, at the focus of X-ray Telescopes (4 systems) but with significantly higher planned effective area than IXPE. Also, PFA will use a new readout ASIC for its gas pixel detector which will result in a much faster time response than IXPE. Please see chapter 15 for a description of the eXTP mission.

\subsubsection{Time Projection Chamber on GEMS}

To overcome the quantum efficiency limitation of GPDs, the time projection chamber (TPC)  was introduced for polarimetry. The TPC essentially turns the GPD on its side, decoupling the drift distance (electrons now drift to the side of the detector) from the absorption depth, which can now be made quite large. The photoelectron track is reconstructed using signals from strip electrodes. In one dimension, normal to the strip direction, the strip number gives the x location of the charges. In the other direction, the drift direction, differences in arrival times give the y co-ordinate of each portion of the track. These two sets of data are combined to reproduce the full photoelectron track. The time projection chambers built for the cancelled NASA GEMS polarimetry mission had absorption depths of $\sim 30$~cm and were filled with 190~torr of dimethyl ether, giving significant quantum efficiency, $>$ 50$\%$ at 3~keV \cite{Hill2012}.

One challenge with the TPC is that it utilizes two different mechanisms to reconstruct the track image. This means that the drift velocity must be very accurately known to convert the time signals to spatial information, and associated with this, that the electric fields must be very uniform. To address this, the GEM detectors used a pulsed x-ray tube to enable in-flight drift velocity monitoring.

One drawback of the TPC is that it gives absolute position information only in the dimension perpendicular to the drift. In the drift/timing direction, only relative positions of electrons in the track are obtained since the start time of the drift is unknown. Thus, the TPC can only provide one-dimensional imaging. Further, when used with a mirror module assembly the large absorption depth defocuses the image, degrading the angular resolution of the system. Nevertheless, because of its quantum efficiency, the devices can provide significant polarization sensitivity in the medium-X-ray energy band.

\subsubsection{The Polarimeter Instrument in X-rays (POLIX)}

POLIX is a modern-day Thomson polarimeter, currently in orbit aboard the Indian X-ray Polarimetry Satellite (XPoSat). POLIX features a central low-Z scatterer (effective area $\sim 400 \, \rm cm^2$) surrounded by four Xenon-filled proportional counters, with an operating energy range from 8-30~keV. XPoSAT is spun around the viewing axis at a fraction of a rotation per minute sampling the source at all azimuthal angles to help reduce  systematic effects in the instrument. For more information on POLIX see Chapter 17 of this volume.

\subsubsection{XL-Calibur}

A good example of an X-ray Compton polarimeter is XL-Calibur (see Chapter 18). This balloon-borne payload features a compact scattering polarimeter at the focus of a large, multilayer-coated X-ray mirror system. The polarimeter consists of an 8-cm-long beryllium scattering rod, surrounded by an array of Cadmium-Zinc-Telluride (CZT) detectors giving $\sim$ 100$\%$ absorption over XL-Calibur’s 15-80 keV operating range. The length of the scattering rod ensures efficient operation, but requires a significant area of absorbing detectors and a resultant significant background. To reduce this, the whole detector assembly is encased in a Bismuth Germanate active shield run in anti-coincidence with the CZT detectors to provide substantial background reduction. XL-Calibur was flown in July 2024 between Sweden and Canada, giving around 1 week of observing time which included $\sim$ 43 hours observing Cyg X-1 \cite{Awaki2025Cyg} and $\sim$ 19 hrs on the Crab \cite{Awaki2025Crab}.

\subsubsection{The Rocket Experiment Demonstration of a Soft-X-ray Polarimeter (REDSoX)}

REDSoX is a Bragg polarimeter currently being developed for launch on a sounding rocket (see Chapter 19). REDSoX operates in the 0.2 – 0.4 keV band and uses laterally graded multilayer coatings to provide the polarization sensitivity. As with crystals, multilayers can give constructive interference at specific angles, following the Bragg condition. By using critical angle transmission gratings, just below an X-ray telescope, to spectrally disperse the beam onto a multilayer-coated reflector which is graded laterally to provide the correct $d$ spacing for the portion of the spectrum that is incident at that point, the reflected flux over the REDSoX band is efficiently directed on to CCD detectors. By orienting 3 sets of gratings plus multilayer reflector plus CCD at $120^{\circ}$ intervals around the axis of the incident X-rays, sufficient azimuthal sampling is provided to enable polarization measurements. REDSoX is scheduled for launch around 2028, and its prime target is isolated neutron stars whose output peaks in the ultra-soft REDSoX energy band. A satellite version, GoSoX is also under study. See Chapter~19 for details.

\subsubsection{The Compton Spectrometer and Imager (COSI)}

COSI is a NASA Small Explorer mission slated for launch in 2027, and while strictly speaking a gamma ray instrument it offers polarization sensitivity in the hard-x-ray region \cite{Tomsick2024}. The COSI payload is a 2 x 2 array of germanium crossed strip detectors, stacked 4 deep. Each strip detector has 64 readout strips, 32 each side, with the rear strips orthogonal to the front. In this way, the 2-D location of an event is given by the intersection of the upper and lower strips triggered by the event. The third co-ordinate, the depth of the interaction within the germanium crystal, is given by the difference in signal arrival time between the front and rear electrodes. The use of germanium, which must be kept at cryogenic temperatures ensures very high spectral resolution, vital for emission line measurements, but which also translates into more accurate Compton kinematics and hence more precise angular resolution. The detector operates as a true 3-D pixelated device, capable of measuring polarization from AGN, Galactic black holes and from gamma ray bursts.


\section{Outlook}
\label{sec:outlook}

After a hiatus of 50 years, X-ray polarimetry has emerged as an essential tool for high energy astrophysics. New missions, driven by advances in instrumentation, have significantly advanced our understanding of many classes of high energy astrophysical sources. Deeper theoretical modeling is needed to fully understand the results obtained to date. The results have also raised many new questions. Moving forward, new observatories that extend the energy band and sensitivity of X-ray polarimetry will be needed to answer our current questions and evoke new ones. The publication of this volume is well timed to summarize the first results obtained from X-ray polarimetry covering a broad range of source classes and highlight new questions to be addressed by continued theoretical work and missions to be launched in the near future.


\begin{acknowledgement}

We thank both the Instituto Nazionale di Astrofisica (INAF) and the Instituto Nazionale di Fisica Nucleare (INFN) for producing Figures~\ref{fig:GPD} and \ref{fig:Track}. We thank Andrea Gnarini for comments which improved the chapter.

\end{acknowledgement}

\bibliographystyle{utphys}
\bibliography{ref.bib}

\providecommand{\href}[2]{#2}\begingroup\raggedright\begin{thebibliography}{10}

\bibitem{Novick1972}
R.~{Novick}, M.~C. {Weisskopf}, R.~{Berthelsdorf}, R.~{Linke}, and R.~S.
  {Wolff}, ``{Detection of X-Ray Polarization of the Crab Nebula},''
  \href{https://dx.doi.org/10.1086/180938}{{\em \apjl} {\bfseries 174} (May,
  1972) L1}.

\bibitem{Weisskopf1978}
M.~C. {Weisskopf}, E.~H. {Silver}, H.~L. {Kestenbaum}, K.~S. {Long}, and
  R.~{Novick}, ``{A precision measurement of the X-ray polarization of the Crab
  Nebula without pulsar contamination.},''
  \href{https://dx.doi.org/10.1086/182648}{{\em \apjl} {\bfseries 220} (Mar.,
  1978) L117--L121}.

\bibitem{KaaretWSPC}
P.~{Kaaret}, \href{https://dx.doi.org/10.1142/9789811203800_0014}{``{X-ray
  Polarimetry},''} in {\em The WSPC Handbook of Astronomical Instrumentation,
  Volume 4: X-Ray Astronomical Instrumentation,}, D.~N. {Burrows}, ed.,
  pp.~281--300.
\newblock World Scientific, 2021.

\bibitem{Rybicki1986}
G.~B. {Rybicki} and A.~P. {Lightman}, {\em {Radiative Processes in
  Astrophysics}}.
\newblock Wiley, 1986.

\bibitem{Ursini2023}
F.~{Ursini}, A.~{Marinucci}, {\em et~al.}, ``{Mapping the circumnuclear regions
  of the Circinus galaxy with the Imaging X-ray Polarimetry Explorer},''
  \href{https://dx.doi.org/10.1093/mnras/stac3189}{{\em \mnras} {\bfseries 519}
  no.~1, (Feb., 2023) 50--58},
  \href{https://arxiv.org/abs/2211.01697}{{\ttfamily arXiv:2211.01697
  [astro-ph.HE]}}.

\bibitem{Veledina2024}
A.~{Veledina}, F.~{Muleri}, {\em et~al.}, ``{Cygnus X-3 revealed as a Galactic
  ultraluminous X-ray source by IXPE},''
  \href{https://dx.doi.org/10.1038/s41550-024-02294-9}{{\em Nature Astronomy}
  {\bfseries 8} (Aug., 2024) 1031--1046},
  \href{https://arxiv.org/abs/2303.01174}{{\ttfamily arXiv:2303.01174
  [astro-ph.HE]}}.

\bibitem{Marin2023}
F.~{Marin}, E.~{Churazov}, {\em et~al.}, ``{X-ray polarization evidence for a
  200-year-old flare of Sgr A$^{*}$},''
  \href{https://dx.doi.org/10.1038/s41586-023-06064-x}{{\em \nat} {\bfseries
  619} no.~7968, (July, 2023) 41--45},
  \href{https://arxiv.org/abs/2304.06967}{{\ttfamily arXiv:2304.06967
  [astro-ph.HE]}}.

\bibitem{Sunyaev1985}
R.~A. {Sunyaev} and L.~G. {Titarchuk}, ``{Comptonization of low-frequency
  radiation in accretion disks Angular distribution and polarization of hard
  radiation},'' {\em \aap} {\bfseries 143} no.~2, (Feb., 1985) 374--388.

\bibitem{Chandrasekhar1946}
S.~{Chandrasekhar}, ``{On the Radiative Equilibrium of a Stellar Atmosphere.
  X.},'' \href{https://dx.doi.org/10.1086/144816}{{\em \apj} {\bfseries 103}
  (May, 1946) 351}.

\bibitem{Angel1969}
J.~R.~P. {Angel}, ``{Polarization of Thermal X-Ray Sources},''
  \href{https://dx.doi.org/10.1086/150185}{{\em \apj} {\bfseries 158} (Oct.,
  1969) 219}.

\bibitem{Poutanen1996}
J.~{Poutanen} and R.~{Svensson}, ``{The Two-Phase Pair Corona Model for Active
  Galactic Nuclei and X-Ray Binaries: How to Obtain Exact Solutions},''
  \href{https://dx.doi.org/10.1086/177865}{{\em \apj} {\bfseries 470} (Oct.,
  1996) 249}, \href{https://arxiv.org/abs/astro-ph/9605073}{{\ttfamily
  arXiv:astro-ph/9605073 [astro-ph]}}.

\bibitem{Schnittman2009}
J.~D. {Schnittman} and J.~H. {Krolik}, ``{X-ray Polarization from Accreting
  Black Holes: The Thermal State},''
  \href{https://dx.doi.org/10.1088/0004-637X/701/2/1175}{{\em \apj} {\bfseries
  701} no.~2, (Aug., 2009) 1175--1187},
  \href{https://arxiv.org/abs/0902.3982}{{\ttfamily arXiv:0902.3982
  [astro-ph.HE]}}.

\bibitem{Connors1980}
P.~A. {Connors}, T.~{Piran}, and R.~F. {Stark}, ``{Polarization features of
  X-ray radiation emitted near black holes.},''
  \href{https://dx.doi.org/10.1086/157627}{{\em \apj} {\bfseries 235} (Jan.,
  1980) 224--244}.

\bibitem{Matt1993}
G.~{Matt}, ``{X-ray polarization properties of a centrally illuminated
  accretion disc.},'' \href{https://dx.doi.org/10.1093/mnras/260.3.663}{{\em
  \mnras} {\bfseries 260} (Feb., 1993) 663--674}.

\bibitem{Burn1966}
B.~J. {Burn}, ``{On the depolarization of discrete radio sources by Faraday
  dispersion},'' \href{https://dx.doi.org/10.1093/mnras/133.1.67}{{\em \mnras}
  {\bfseries 133} (Jan., 1966) 67}.

\bibitem{Heisenberg1936}
W.~{Heisenberg} and H.~{Euler}, ``{Folgerungen aus der Diracschen Theorie des
  Positrons},'' \href{https://dx.doi.org/10.1007/BF01343663}{{\em Zeitschrift
  fur Physik} {\bfseries 98} no.~11-12, (Nov., 1936) 714--732}.

\bibitem{Adler1971}
S.~L. {Adler}, ``{Photon splitting and photon dispersion in a strong magnetic
  field.},'' \href{https://dx.doi.org/10.1016/0003-4916(71)90154-0}{{\em Annals
  of Physics} {\bfseries 67} (Jan., 1971) 599--647}.

\bibitem{Heyl1997}
J.~S. {Heyl} and L.~{Hernquist}, ``{Birefringence and dichroism of the QED
  vacuum},'' \href{https://dx.doi.org/10.1088/0305-4470/30/18/022}{{\em Journal
  of Physics A Mathematical General} {\bfseries 30} no.~18, (Sept., 1997)
  6485--6492}, \href{https://arxiv.org/abs/hep-ph/9705367}{{\ttfamily
  arXiv:hep-ph/9705367 [astro-ph]}}.

\bibitem{Heyl2002}
J.~S. {Heyl} and N.~J. {Shaviv}, ``{QED and the high polarization of the
  thermal radiation from neutron stars},''
  \href{https://dx.doi.org/10.1103/PhysRevD.66.023002}{{\em \prd} {\bfseries
  66} no.~2, (July, 2002) 023002},
  \href{https://arxiv.org/abs/astro-ph/0203058}{{\ttfamily
  arXiv:astro-ph/0203058 [astro-ph]}}.

\bibitem{Costa2024}
E.~{Costa}, ``{Scattering Polarimetry in the Hard X-ray Range},''
  \href{https://dx.doi.org/10.3390/instruments8010020}{{\em Instruments}
  {\bfseries 8} no.~1, (Mar., 2024) 20}.

\bibitem{Friis2018}
M.~{Friis}, M.~{Kiss}, V.~{Mikhalev}, M.~{Pearce}, and H.~{Takahashi}, ``{The
  PoGO+ Balloon-Borne Hard X-ray Polarimetry Mission},''
  \href{https://dx.doi.org/10.3390/galaxies6010030}{{\em Galaxies} {\bfseries
  6} no.~1, (Mar., 2018) 30},
  \href{https://arxiv.org/abs/1803.02106}{{\ttfamily arXiv:1803.02106
  [astro-ph.IM]}}.

\bibitem{McConnell2009}
M.~L. {McConnell}, C.~{Bancroft}, P.~F. {Bloser}, T.~{Connor}, J.~{Legere}, and
  J.~M. {Ryan}, \href{https://dx.doi.org/10.1117/12.826407}{``{GRAPE: a
  balloon-borne gamma-ray polarimeter},''} in {\em UV, X-Ray, and Gamma-Ray
  Space Instrumentation for Astronomy XVI}, O.~H. {Siegmund}, ed., vol.~7435 of
  {\em Society of Photo-Optical Instrumentation Engineers (SPIE) Conference
  Series}, p.~74350J.
\newblock Aug., 2009.

\bibitem{Forot2008}
M.~{Forot}, P.~{Laurent}, I.~A. {Grenier}, C.~{Gouiff{\`e}s}, and F.~{Lebrun},
  ``{Polarization of the Crab Pulsar and Nebula as Observed by the
  INTEGRAL/IBIS Telescope},'' \href{https://dx.doi.org/10.1086/593974}{{\em
  \apjl} {\bfseries 688} no.~1, (Nov., 2008) L29},
  \href{https://arxiv.org/abs/0809.1292}{{\ttfamily arXiv:0809.1292
  [astro-ph]}}.

\bibitem{Chattopadhyay2021}
T.~{Chattopadhyay}, ``{Hard X-ray polarimetry{\textemdash}an overview of the
  method, science drivers, and recent findings},''
  \href{https://dx.doi.org/10.1007/s12036-021-09769-5}{{\em Journal of
  Astrophysics and Astronomy} {\bfseries 42} no.~2, (Oct., 2021) 106},
  \href{https://arxiv.org/abs/2104.05244}{{\ttfamily arXiv:2104.05244
  [astro-ph.HE]}}.

\bibitem{Kaaret1990}
P.~{Kaaret}, R.~{Novick}, C.~{Martin}, P.~{Shaw}, T.~{Hamilton}, R.~{Suniaev},
  I.~{Lapshov}, E.~{Silver}, M.~{Weisskopf}, and R.~{Elsner}, ``{The Stellar
  X-ray Polarimeter - A focal plane polarimeter for the Spectrum X-Gamma
  mission},'' \href{https://dx.doi.org/10.1117/12.55660}{{\em Optical
  Engineering} {\bfseries 29} (July, 1990) 773--780}.

\bibitem{Costa2001}
E.~{Costa}, P.~{Soffitta}, R.~{Bellazzini}, A.~{Brez}, N.~{Lumb}, and
  G.~{Spandre}, ``{An efficient photoelectric X-ray polarimeter for the study
  of black holes and neutron stars},''
  \href{https://dx.doi.org/10.1038/35079508}{{\em \nat} {\bfseries 411}
  no.~6838, (June, 2001) 662--665},
  \href{https://arxiv.org/abs/astro-ph/0107486}{{\ttfamily
  arXiv:astro-ph/0107486 [astro-ph]}}.

\bibitem{Feng2019}
H.~{Feng}, W.~{Jiang}, {\em et~al.}, ``{PolarLight: a CubeSat X-ray polarimeter
  based on the gas pixel detector},''
  \href{https://dx.doi.org/10.1007/s10686-019-09625-z}{{\em Experimental
  Astronomy} {\bfseries 47} no.~1-2, (Apr., 2019) 225--243},
  \href{https://arxiv.org/abs/1903.01619}{{\ttfamily arXiv:1903.01619
  [astro-ph.IM]}}.

\bibitem{Kitaguchi2019}
T.~{Kitaguchi}, K.~{Black}, T.~{Enoto}, A.~{Hayato}, J.~E. {Hill}, W.~B.
  {Iwakiri}, P.~{Kaaret}, T.~{Mizuno}, and T.~{Tamagawa}, ``{A convolutional
  neural network approach for reconstructing polarization information of
  photoelectric X-ray polarimeters},''
  \href{https://dx.doi.org/10.1016/j.nima.2019.162389}{{\em Nuclear Instruments
  and Methods in Physics Research A} {\bfseries 942} (Oct., 2019) 162389},
  \href{https://arxiv.org/abs/1907.06442}{{\ttfamily arXiv:1907.06442
  [astro-ph.IM]}}.

\bibitem{Peirson2021}
A.~L. {Peirson}, R.~W. {Romani}, H.~L. {Marshall}, J.~F. {Steiner}, and
  L.~{Baldini}, ``{Deep ensemble analysis for Imaging X-ray Polarimetry},''
  \href{https://dx.doi.org/10.1016/j.nima.2020.164740}{{\em Nuclear Instruments
  and Methods in Physics Research A} {\bfseries 986} (Jan., 2021) 164740},
  \href{https://arxiv.org/abs/2007.03828}{{\ttfamily arXiv:2007.03828
  [astro-ph.IM]}}.

\bibitem{Hill2012}
J.~E. {Hill}, R.~G. {Baker}, {\em et~al.},
  \href{https://dx.doi.org/10.1117/12.928435}{``{The design and qualification
  of the GEMS x-ray polarimeters},''} in {\em Space Telescopes and
  Instrumentation 2012: Ultraviolet to Gamma Ray}, T.~{Takahashi}, S.~S.
  {Murray}, and J.-W.~A. {den Herder}, eds., vol.~8443 of {\em Society of
  Photo-Optical Instrumentation Engineers (SPIE) Conference Series}, p.~84431Q.
\newblock Sept., 2012.

\bibitem{Awaki2025Cyg}
H.~{Awaki}, M.~G. {Baring}, {\em et~al.}, ``{XL-Calibur Polarimetry of Cyg X-1
  Further Constrains the Origin of Its Hard-state X-Ray Emission},''
  \href{https://dx.doi.org/10.3847/1538-4357/ae0f1d}{{\em \apj} {\bfseries 994}
  no.~1, (Nov., 2025) 37}, \href{https://arxiv.org/abs/2507.23126}{{\ttfamily
  arXiv:2507.23126 [astro-ph.HE]}}.

\bibitem{Awaki2025Crab}
H.~{Awaki}, M.~G. {Baring}, {\em et~al.}, ``{XL-Calibur measurements of
  polarized hard X-ray emission from the Crab},''
  \href{https://dx.doi.org/10.1093/mnrasl/slaf026}{{\em \mnras} {\bfseries 540}
  no.~1, (June, 2025) L34--L40},
  \href{https://arxiv.org/abs/2503.14307}{{\ttfamily arXiv:2503.14307
  [astro-ph.HE]}}.

\bibitem{Tomsick2024}
J.~{Tomsick}, S.~{Boggs}, {\em et~al.},
  \href{https://dx.doi.org/10.22323/1.444.0745}{``{The Compton Spectrometer and
  Imager},''} in {\em 38th International Cosmic Ray Conference}, p.~745.
\newblock Sept., 2024.
\newblock \href{https://arxiv.org/abs/2308.12362}{{\ttfamily arXiv:2308.12362
  [astro-ph.HE]}}.

\end{thebibliography}\endgroup

\end{document}